\documentclass{kapproc} 
\usepackage{graphics}

\kluwerbib

\begin{document}

\articletitle{3D modeling of 1612 MHz OH masers}

\articlesubtitle{Monte Carlo modeling of the maser shells and the
  amplified stellar image}

\author{W.H.T. Vlemmings}
\affil{Sterrewacht Leiden\\
Niels Bohr weg 2, 2300 RA Leiden, The Netherlands}
\email{vlemming@strw.leidenuniv.nl}

\begin{abstract}
 We present the first results of our 3D Monte Carlo maser radiative
 transfer code, used to model the 1612 MHz OH maser shell and the
 amplification of emission from the stellar radio-photosphere.
\end{abstract}


\section*{Introduction}

In the circumstellar envelope of AGB stars, both the main line (1665
and 1667 MHz) and the satellite line (1612 MHz) OH maser transitions
have been observed extensively. The 1612 MHz masers occur at~$\approx
10^{16}$~cm from the central star.

The 1612 MHz maser line profile often shows a characteristic double
peaked spectrum that is assumed to be the signature of an expanding
shell (e.g. Reid et al. 1977). The exact shape however, is determined
by local conditions in the maser shell and the level of saturation.

Recently it has been suggested that emission from the radio-sphere of
the star will be amplified by the maser screen in front of the
star. This amplification will cause a bright maser spots with several
observable characteristics (Vlemmings, 2000).  Strong evidence for
this theory was presented recently when the most blue-shifted 1667 MHz
maser spot of U~Her was found to coincide with the Hipparcos optical
position to within 15 mas (van Langevelde et al., 2000 \& these
proceedings)

Modeling the various observed shapes of the maser profile by, and
examining the effect of a background star are the main goals of this
work.

\section{Modeling \& Results}

Maser emission is highly anisotropic as a result of beaming due to the
maser geometry and the structure of velocity coherent paths through
the medium. Additional anisotropy can be caused by density or
pumping inhomogeneities and the contributions of the background
source. This means that full 3-dimensional modeling is essential to
accurately describe the maser.

Here we have used a fully 3D Monte Carlo code that solves the 1612 MHz
OH maser in a way similar to the 2D method used by Spaans
\&~van~Langevelde (1992).

The first results are presented in Fig.1 for a typical OH shell with
and without a background star. The expansion velocity is $10$~km/s and
the OH number density is $4$~cm$^{-3}$ throughout the maser shell. As
can be seen, the effect of a background star on the OH maser spectrum
is small, while the high resolution channel map clearly shows the high
brightness stellar image.

\begin{figure*}
\resizebox{\hsize}{!}{\rotatebox{-90}{\includegraphics{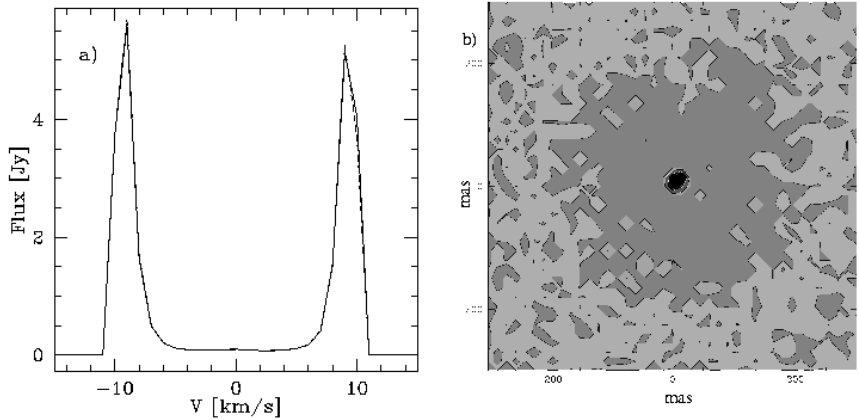}}}
\caption{a) Modeled OH maser spectrum with (solid) and without (dashed)
  background star. b) Map of the most blue-shifted velocity channel
  for the model with a background star. Black is the highest intensity.}
\end{figure*}

\end{document}